\documentclass[aps,pre,showpacs]{revtex4}
\usepackage[dvips]{graphicx,color} 
\usepackage{float,times,amsbsy}
\usepackage{amssymb}
\usepackage{booktabs}
\usepackage{calc}
\usepackage{comment}
\usepackage{bm}
\usepackage[latin1]{inputenc}
\usepackage{amsmath,amssymb} 
\usepackage{epstopdf}
\usepackage{verbatim}
\usepackage{calligra}
\DeclareMathAlphabet{\mathcalligra}{T1}{calligra}{m}{n}
\DeclareFontShape{T1}{calligra}{m}{n}{<->s*[2.2]callig15}{}
\newcommand{\scrs}{\ensuremath{\mathcalligra{s}}}
\newcommand{\ra}[1]{\renewcommand{\arraystretch}{#1}}
\begin{document}

\title{``Cooling by heating'' -- demonstrating the significance of the
longitudinal specific heat}
\author{Jon J. Papini, Jeppe C. Dyre, and Tage Christensen}\email{tec@ruc.dk}
\affiliation{DNRF Centre ``Glass and Time'', IMFUFA, Department of Sciences,
Roskilde University,
Postbox 260, DK-4000 Roskilde, Denmark}
\date{\today}

\begin{abstract}
Heating a solid sphere at the surface induces mechanical stresses inside the sphere. If a finite amount of heat is supplied, the stresses gradually disappear as temperature becomes homogeneous throughout the sphere. We show that before this happens, there is a temporary lowering of pressure and density in the interior of the sphere, inducing a transient lowering of the temperature here. For ordinary solids this effect is small because $c_p\cong c_V$. For fluent liquids the effect is negligible because their dynamic shear modulus vanishes. For a liquid at its glass transition, however, the effect is generally considerably larger than in solids. This paper presents analytical solutions of the relevant coupled thermoviscoelastic equations. In general, there is a difference between the isobaric specific heat, $c_p$, measured at constant isotropic pressure and the longitudinal specific heat, $c_l$, pertaining to mechanical boundary conditions that confine the associated expansion to be longitudinal. In the exact treatment of heat propagation the heat diffusion constant contains $c_l$ rather than $c_p$. We show that the key parameter controlling the magnitude of the ''cooling-by-heating`` effect is the relative difference between these two specific heats. For a typical glass-forming liquid, when temperature at the surface is increased by 1 K, a lowering of the temperature in the sphere center of order 5 mK is expected if the experiment is performed at the glass transition. The cooling-by-heating effect is confirmed by measurements on a 19 mm diameter glucose sphere at the glass transition.
\end{abstract}

\maketitle

\section{Introduction}

Most solids and liquids expand when heated. Heat diffusion is a notoriously slow process, and heating a solid sample at its surface induces stresses in the sample that only disappear when temperature gradually becomes again homogeneous throughout. Heating a lightly fluent fluid that has a free surface (i.e., is free to expand), on the other hand, makes the entire sample expand on the time scale set by the sound velocity and sample dimensions. In this case there are no transient stresses beyond the acoustic time scale. A liquid close to its glass transition provides an interesting case in
between solid and fluid behavior. Such a liquid behaves like a solid on time scales shorter than the Maxwell relaxation time $\tau_\text{M}=\eta/G_\infty$ where $\eta$ is the shear viscosity and $G_\infty$ the instantaneous shear modulus. The Maxwell relaxation time becomes longer than one second when a liquid approaches its calorimetric glass transition, implying that induced stresses survive for seconds or more. 

This paper discusses the ``cooling-by-heating'' effect that arises when a sample is heated at a free surface. We show that this effect, which is present in all hard solids with a non-zero thermal expansion coefficient, is generally magnified considerably for glass-forming liquids close to their
glass-transition temperature $T_g$. This is because close to $T_g$ the liquid is solid-like by having a large, non-zero dynamic shear modulus on short time scales and, at the same time, is liquid-like by having a fairly large thermal expansion coefficient.

Returning to the case of a solid, what happens when heat is supplied at the (free) surface of a spherical sample? The outermost layers {\it attempt} to expand, obviously, but {\it a priori} one may imagine two different possibilities: 1) the expansion presses inwards, resulting in an increase of the pressure at the center of the sphere; or: 2) the expansion turns outwards, thus transmitting a negative pressure into the sphere. Which of the two possibilities that applies is answered by the application of standard thermo-elasticity theory to the problem of calculating the stresses induced by the heating. This is done in the present paper. It turns out that case 2 applies---the sphere expands and pressure decreases in the interior of the sphere. This induces an adiabatic cooling inside the sphere. The phenomenon of cooling caused by heating at the surface is referred to below as the cooling-by-heating effect.

The solution of the coupled thermomechanical equations detailed in Sec. \ref{sec2} shows that the cooling-by-heating effect is proportional to the difference between the reciprocals of the isobaric specific heat, $c_p$ and the longitudinal
specific heat, $c_l$ (all specific heats are per unit volume); the latter quantity was introduced and discussed in Refs. \onlinecite{Tage_1,Tage_2}.
The longitudinal specific heat is related to the isochoric specific heat, $c_V$, by
\begin{equation}\label{defcl}
 c_l\,=\,\frac{M_S}{M_T} c_V\, ,
\end{equation}
where $M_S$ and $M_T$ are the adiabatic and isothermal longitudinal moduli respectively. This is analogous to the standard thermodynamic relation $c_p=(K_S/K_T)c_V$ relating the isobaric specific heat to the isochoric specific heats in terms of the adiabatic, $K_S$ and isothermal, $K_T$ bulk moduli. Since $M_S=K_S+(4/3)G$ and $M_T=K_T+(4/3)G$, where G is shear modulus, one readily finds that $c_l$ is in between $c_V$ and $c_p$. As we shall see, the relative difference $a_l=(c_p-c_l)/c_p$ controls the strength of the cooling-by-heating effect, and we thus term this quantity the ``longitudinal thermomechanical coupling constant''. Combining the equations above $a_l$ is found to be the product of two factors \cite{Tage_1},
\begin{equation}\label{a_l}
a_l\equiv \frac{c_p-c_l}{c_p}
\,=\,\frac{4}{3}\frac{G}{M_T}\frac{c_p-c_V}{c_p}.
\end{equation}
The first factor shows that there is only cooling by heating if the shear modulus is
non-vanishing compared to the longitudinal modulus. The other factor, ``the thermomechanical coupling'', $a=(c_p-c_V)/c_p$ is a dimensionless measure of the coupling between thermal and mechanical perturbations. It can be expressed in terms of the expansivity, $\alpha_p\equiv (1/V) (\partial V/\partial T)_p$, as follows:
\begin{equation}\label{tmecoup}
a\equiv \frac{c_p-c_V}{c_p}\,=\,\frac{T_0\alpha_p^2 K_T }{c_p}\,,
\end{equation}
where $T_0$ is the temperature. It follows that the cooling-by-heating effect is quadratic in the thermal expansion coefficient $\alpha_p$. 

Since solids typically expand significantly less upon heating than do liquids, the cooling-by-heating effect is generally small in solids. As an example, for solid glucose the thermal expansion coefficient \cite{Parks1928} is $1.1 \cdot 10^{-4}$ K$^{-1}$ close to the glass transition whereas for liquid glucose it is  $3.7 \cdot 10^{-4}$ K$^{-1}$in the same temperature region. This potentially enhances the cooling-by-heating effect by a factor of $11$. However the changes in $c_p$ \cite{Parks1928} from $1.91 \cdot 10^6$ JK$^{-1}$m$^{-3}$  to  $3.05 \cdot 10^6$ JK$^{-1}$m$^{-3}$  and in $K_T$ \cite{DaviesJones53} from $10.75 \cdot 10^9$ Pa to $6.49 \cdot 10^9$ Pa reduces this to a factor of $8$. Here we have used a density of $1.52 \cdot 10^3$ kg m$^{-3}$ to convert specific heat data from mass to volume. It is, however, not unusual for liquid expansivities to be near $10^{-3}$ K$^{-1}$ for which we would expect an enhancement of the thermomechanical coupling $a=(c_p-c_V)/c_p$ by a factor of $30$. The shear modulus of glucose  in the glassy state is $G_\infty=3.1 \cdot 10^9$ Pa as deduced from the shear compliance data of Meyer and Ferry \cite{Meyer65}.

\begin{figure}
\begin{center}
\includegraphics[width=8.6cm]{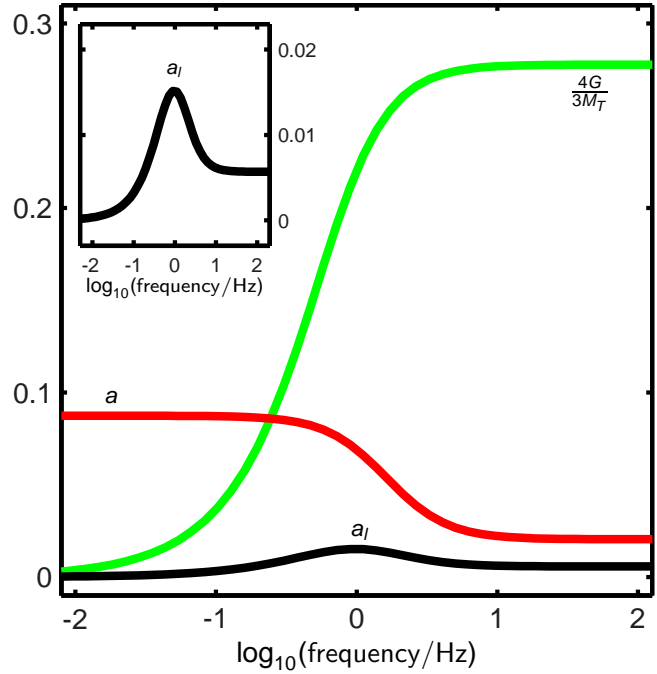}
\caption{\label{fig1} Sketch of overlapping relaxations of $4G/(3M_T)$ (green) and the thermomechanical coupling, $a=T\alpha_p^2K_T/c_p$ (red). The longitudinal coupling constant $a_l=(c_l-c_p)/c_p$ (black) is the product of those two frequency dependent functions. Near the crossing of the two curves the difference between $c_l$ and $c_p$ is generally largest. This determines the time scale of experiments on glass-forming liquids, where the cooling-by-heating effect is particularly large.}
\end{center}
\end{figure}

The above relations all generalize to deal with complex, frequency-dependent (dynamic) specific
heats and moduli and expansivity, which are the relevant quantities when studying glass-forming liquids. Near the
glass transition the cooling-by-heating effect may be studied on second time scales. Here, upon
increasing the frequency, the factor $G/M_T$ of Eq. (\ref{a_l}) increases while at the same time the
factor $(c_p-c_v)/c_p$ decreases. The enhancement of the cooling-by-heating effect is thus
critically dependent on the relative time scales of the different relaxation processes at the glass
transition. If the shear stress relaxes faster than the the volume processes, the cooling-by-heating
effect may not be pronounced. This situation is illustrated in Fig. \ref{fig1}. The model describing
the relaxation behavior between high- and low-frequency values is described in section \ref{sec4}.

The present work discusses the basis of cooling by heating by referring to the equations of standard linear thermoviscoelasticity. Section \ref{sec1} introduces the general framework of thermoelasticity and thermoviscoelasticity. It is shown that the heat diffusion constant involves the longitudinal specific heat. Section \ref{sec2} discusses the case when a finite amount of heat is fed into a sample at its surface at $t=0$, as well as the experimentally easier realized case when temperature is suddenly increased at the surface. That section also presents analytical calculations of the ordinary solid case for which the constitutive properties do not undergo relaxation. Section \ref{sec4} gives calculations of a model glass-forming liquid, i.e., the case when the constitutive properties are frequency dependent. We estimate the effect to be of order $5$ mK in the center of a sphere for a temperature increase at the sphere surface of one Kelvin. Section \ref{sec5} confirms this prediction for measurements on a glucose sphere. Sections VI and \ref{sec6} briefly discuss and summarize the paper.

\section{Thermoelasticity and heat diffusion.}\label{sec1}

Thermoelasticity deals with problems where displacement field $\textbf{u}(\textbf{r},t)$ and
temperature field $T(\textbf{r},t)$ couple. It is a linear theory of small deformations given in
terms of the strain tensor $\epsilon_{ij}=\frac{1}{2}(\frac{\partial u_i}{\partial
x_j}+\frac{\partial u_j}{\partial x_i})$ and small temperature increments $\delta T=T-T_0$ relative
to a reference temperature $T_0$. The material properties of a thermoelastic medium is given by the
linear constitutive equations that expresses stress $\sigma_{ij}$ and increments in entropy density
$\delta \scrs$ in terms of $\epsilon_{ij}$ and $\delta T$. The hydrostatic pressure
is related to the trace of the stress tensor $p= -1/3 \ \sum_i \sigma_{ii}$ and the relative
compression is the trace of the strain tensor $\epsilon= \sum_i \epsilon_{ii}=\boldsymbol{\nabla}
\cdot \textbf{u}$. The following constitutive equations \cite{Christensen} define the shear modulus $G$, the isothermal
bulk modulus, $K_T$, the isochoric specific heat, $c_V$ and the isochoric pressure coefficient
$\beta_V$:
\begin{eqnarray}
 \sigma_{ij}+p \delta_{ij}=2 G (\epsilon_{ij}-\frac{1}{3} \epsilon \delta_{ij}) \label{eq:tve1} \\
  p= -K_T \epsilon + \beta_V \delta T \label{eq:tve2} \\
 \delta \scrs=\beta_V \epsilon +\frac{c_V}{T_0} \delta T\,. \label{eq:tve3}
\end{eqnarray}
We follow Biot \cite{Biot56} in assigning the symbol $\beta$ to the thermodynamic pressure coefficient
\begin{equation}
 \beta_V=\left(\frac{\partial p}{\partial T}\right)_V =\left(\frac{\partial S}{\partial V}\right)_T=
\alpha_p K_T\,.
\end{equation}
The material is furthermore characterized by the heat conductivity, $\lambda$, which enters Fourier's law for the entropy current density $\textbf{j}_\scrs$:
\begin{equation} \label{eq:Fourier}
 \textbf{j}_\scrs=-\frac{\lambda}{T_0} \boldsymbol{\nabla} T\,.
\end{equation}

The interest in thermoelastic problems has since Duhamel \cite{Duhamel1837} mostly been focused on the calculation of thermal stresses deriving from an evolving temperature field. In the classical thermoelasticity theory the displacement field and temperature fields are partially decoupled \cite{Nowacki,Parkus}. This comes from assuming that the development of the temperature can be found independently of the stresses by the conventional heat-diffusion equation:
\begin{equation} \label{eq:chd}
 \frac{\partial \delta T}{\partial t} = D \nabla^2 \delta T\,.
\end{equation}
Here $D$ is a heat diffusion constant. After solving this equation the displacement field can be found from the quasi-static stress equilibrium equation:
\begin{equation} \label{eq:stressequilibrium}
 M_T \boldsymbol{\nabla}(\boldsymbol{\nabla} \cdot
\textbf{u}) -G \boldsymbol{\nabla} \times (\boldsymbol{\nabla} \times \textbf{u}) - \beta_V \boldsymbol{\nabla} \delta T = \textbf{0}
\end{equation}
This approximate theory is referred to as the theory of thermal stresses \cite{Nowacki}. According to many authors  \cite{HetnarskiEslami,Chadwick60,Sneddon72,Nowacki} the correct treatment appeared remarkably late in the development of thermoelastic theory with Biot's paper \cite{Biot56} in 1956. Lessen \cite{Lessen56} considered similar problems the same year. The heat diffusion equation Eq. (\ref{eq:chd}) is replaced by
\begin{equation} \label{eq:entropycons}
 c_V\frac{\partial \delta T}{\partial t} + T_0 \beta_V \frac{\partial \boldsymbol{\nabla} \cdot
\textbf{u}}{\partial t}= \lambda \nabla^2 \delta T\,,
\end{equation}
which follows from entropy conservation
\begin{equation}\label{eq:entropycons2}
 \frac{\partial \scrs}{\partial t}=-\boldsymbol{\nabla} \cdot {\boldsymbol j}_\scrs
\end{equation}
when this is combined with Eqs. (\ref{eq:tve3}) and (\ref{eq:Fourier}). Entropy conservation may seem strange at first sight, but the entropy production per volume associated with heat conduction is $-\textbf{j}_\scrs \cdot \frac{1}{T_0} \boldsymbol{\nabla T} = \frac{\lambda}{T^2_0} (\boldsymbol{\nabla} T)^2$, i.e., a second-order effect disappearing in a linearized theory.

In most cases the ordinary, decoupled heat-diffusion equation is a good approximation in the manner it is used in the theory of thermal stresses. However, this approximate theory is not able to describe the phenomenon of cooling by heating, which is the theme of this paper. It should be noted that the heat-diffusion equation with the diffusion constant containing the isobaric specific heat $c_p$ is exact for the non-viscous liquid state or soft matter with $G=0$, if part of the boundary (with normal vector $\mathbf{n}$) is free to expand, i.e., $\sum_j\sigma_{ij} n_j=0$. The proof runs as follows: The assumption $G/M_T=0$ simplifies Eq. (\ref{eq:stressequilibrium}) to
\begin{equation}
 \boldsymbol{\nabla}(K_T \boldsymbol{\nabla} \cdot \textbf{u}-\beta_V \delta T)= \textbf{0}\,.
\end{equation}
However, the terms under the gradient is according to Eq. (\ref{eq:tve2}) nothing but minus the pressure increment. Thus we conclude this pressure increment is uniform in space and only depends on time. Moreover, Eq. (\ref{eq:tve1}) ensures that all diagonal elements of the stress tensor are identical and equal to minus this pressure increment. If the normal component $\sum_j\sigma_{ij} n_j$ is zero on part of the boundary, it follows that the pressure is also zero there, but then it is zero throughout the body. Equation (\ref{eq:tve2}) then reduces to $ \boldsymbol{\nabla} \cdot \textbf{u}=\alpha_p \delta T$. Inserting this in the entropy equation Eq. (\ref{eq:entropycons}), one arrives at the ordinary decoupled heat diffusion equation with $D_p=\lambda/c_p$
\begin{equation}
  \frac{\partial \delta T}{\partial t} = D_p \nabla^2 \delta T \,,
\end{equation}
when noticing that $c_p=c_V+T_0 \frac{\beta^2_V}{K_T}=c_V+T_0 \alpha^2_p K_T$.

As se  have seen, the temperature field in general does not exactly obey a diffusion equation. It
does so when $c_p=c_V$ ($\Leftrightarrow \beta_V=0$) or for certain boundary conditions when $G=0$.
However, as emphasized by Biot \cite{Biot56}, the entropy density in fact does fulfill a diffusion
equation and moreover with a diffusion constant containing the ubiquitous longitudinal specific heat: Applying the
divergence operator to the inertia-free stress equilibrium equation Eq. (\ref{eq:stressequilibrium}), gives
\begin{equation}
  \nabla^2 \epsilon= \frac{\beta_V}{M_T}  \nabla^2 \delta T\,.
\end{equation}
Applying the Laplacian to the constitutive equation (\ref{eq:tve3}) yields
\begin{equation}
 \nabla^2 \scrs=(\frac{\beta_V^2}{M_T}+\frac{c_V}{T_0}) \nabla^2 \delta T\,.
\end{equation}
Fouriers law and the entropy conservation Eq. (\ref{eq:entropycons2}) gives
\begin{equation}
 \frac{\partial \scrs}{\partial t}= \frac{\lambda}{T_0}\nabla^2  \delta T\,,
\end{equation}
and thus
\begin{equation}
 \frac{\partial \scrs}{\partial t}= \frac{\lambda}{c_l} \nabla^2 \scrs \,,
\end{equation}
with $c_l=c_V+T_0\frac{\beta_V^2}{M_T}$ being the longitudinal specific heat \cite{Tage_1,Tage_2}. 

Note that this result is limited to the inertia-free cases. If one wishes to study coupled mechanical and thermal waves, the inertia-term $\rho \frac{\partial^2}{\partial t^2} \textbf{u}$ must be added on the right side of Eq. (\ref{eq:stressequilibrium}). Solutions of the equations in this case have been studied extensively \cite{Nowacki} also in the spherically symmetric situation. Note, however, that acoustic wavelengths are much longer than thermal wavelengths. Thus for a sample of a certain size there is an interesting time regime where acoustic waves have settled, but thermal diffusion has barely begun. Take as an example a sphere of radius $1$cm. For a typical sound velocity of $10^3$ m/s and heat diffusion constant of $10^{-7}$ m$^2$/s, the sound traveling time is $10$ $\mu$s while the diffusion time is $1$ ks. It is within this time regime we will find the cooling-by-heating phenomenon. Although the solution restricted to the inertia-free case that we present below is in principle contained in the coupled acoustic-thermal wave solutions including inertia, the phenomenon is obscured by the complicated structure of these solutions and seems not to have been recognized.

The thermoelastic theory that was originally developed for elastic solids without relaxation is
easily extended to a thermoviscoelastic theory taking relaxation of all the constitutive
parameters into account, as it is necessary for relaxing liquids near the glass transition. The most
straightforward way of generalizing is to interpret the equations in the frequency domain allowing
all the constitutive parameters to be complex functions of the angular frequency $\omega$. The cases
we study in the frequency domain cover thus both solids and thermoviscoelastic liquids, but can only
be transformed analytically into the time domain for solids. For relaxing liquids one must do the
transformation numerically.

\section{Analytical solutions of the sphere-heating problem}\label{sec2}

\subsection{The case when the heat flow is controlled at a mechanically free boundary}\label{sec2a}

This subsection presents the analytical solution in the frequency domain to the situation when a
sphere of a general viscoelastic material is subjected to a periodic heat input at the
surface. The solution shows the temperature in the center at high frequencies varying $180^\circ$
out-of-phase with respect to the heat oscillation at the surface, indicating the cooling-by-heating
effect. In order to give a more lucid and transparent understanding of the phenomenon we translate
the solution to the time domain. This can be done analytically by an inverse Laplace transformation
if there is no frequency dependence of the constitutive properties. That is, we calculate the
temperature and stress profile throughout the sphere following a heat-step input at the surface at
time zero.

Consider the case when a periodically varying heat $\delta Q(t)={\rm Re}\left\lbrace \delta
Q e^{i\omega t}\right\rbrace $ is supplied at the surface of a sphere of radius $R$. The
surface is assumed to be mechanically non-clamped, i.e., the sphere is free to expand. This
translates into the boundary condition that the radial component of the stress tensor is zero at the
surface, $\sigma_{rr}(R,\omega)=0$. We wish to calculate how the periodically varying temperature
and displacement fields vary throughout the sphere, i.e., to calculate the complex
frequency-dependent amplitudes of temperature, $\delta T(r,\omega)$, and radial displacement field,
$u(r,\omega)$. From these quantities the stress components $\sigma_{rr}(r,\omega)$, etc, may be
calculated.

Denoting the angular frequency by $\omega$, the position vector by $\bf r$, the complex
frequency-dependent radial displacement field by ${\bf u}(r,\omega)= u(r,\omega){\bf r}/r$, the
coupled thermoelastic equations (\ref{eq:stressequilibrium}) and (\ref{eq:entropycons}) become
\begin{eqnarray}\label{eq:stress_eq_sph}
\frac{\partial}{\partial r} \left[
M_T \, r^{-2}\frac{\partial}{\partial r} (r^2  u)-  \beta_V \delta T \right] &=& 0 \\
(i\omega)c_V \delta T+(i\omega)T_0 \beta_V \,r^{-2}\frac{\partial}{\partial r} (r^2 u) &=& \lambda \, r^{-2} \frac{\partial}{\partial r} (r^2 \frac{\partial}{\partial r} \delta T) \,.\label{eq:entropycons_sph}
\end{eqnarray}

The four boundary conditions are: 
\begin{enumerate}
\item No displacement at the center: $u(0,\omega)=0$;
\item No temperature gradient at the center: $\frac{\partial \delta T  }{
\partial r}(0,\omega)=0$;
\item Free surface, i.e., no radial stresses at the surface: $\sigma_{rr}(R,\omega)=0$;
\item Heat supply boundary condition at the surface: $\lambda \frac{\partial \delta T
}{\partial r}(R,\omega)=i\omega\frac{\delta Q }{4\pi R^2}$.
\end{enumerate}

Denote the volume of the sphere by $V_0=\frac{4}{3}\pi R^3$ and define the complex frequency-dependent thermal wavevector by $k=\sqrt{i\omega c_l (\omega)/ \lambda}$. Define furthermore the functions
\begin{eqnarray}
f_1(r/R, k^2 R^2)&=&\frac{1}{3} \frac{\sinh(kr)}{kr} \frac{(kR)^3}{kR \cosh(kR)-\sinh(kR)} \label{eq:f1}\\
f_2(r/R, k^2 R^2)&=&\Bigl(\frac{R}{r}\Bigr)^3 \frac{(kr) \cosh(kr)-\sinh(kr)}{(kR)
\cosh(kR)-\sinh(kR)} \label{eq:f2}\,.
\end{eqnarray}

Introduce the characteristic heat diffusion time
\begin{equation}
 \tau= (c_l (\omega)/ \lambda)R^2. \label{eq:tau}
\end{equation}
Then one has $k^2 R^2=i\omega\tau$ and the solutions to Eqs. (\ref{eq:stress_eq_sph}) and (\ref{eq:entropycons_sph}) are 
\begin{eqnarray}
\delta T(r,\omega)&=&\frac{1}{V_0 c_l}\left\lbrace -a_l
+f_1(r/R, i \omega \tau)   
\right\rbrace\delta Q \label{eq:solution_1}\\
u (r,\omega)&=&\frac{1}{3} \frac{\alpha_p}{V_0 c_p} r  \left\lbrace
\frac{4}{3} \frac{G}{M_S} +\frac{K_S}{M_S} f_2(r/R , i \omega \tau)  \right\rbrace \delta Q \label{eq:solution_2}\,.
\end{eqnarray}

These solutions were found by the transfer-matrix approach (see Ref. \onlinecite{Tage_2} and Appendix \ref{App1}), and can be verified by insertion, noticing that $f_1(\rho,s)=\frac{1}{3 \rho^2}\frac{\partial}{\partial \rho}(\rho^3 f_2(\rho,s))$, $\frac{\partial}{\partial \rho} f_1(\rho,s)=\frac{1}{3} \rho s f_2(\rho,s)$ and $\frac{1}{\rho^2}$ $\frac{\partial}{\partial \rho}\rho^2 \frac{\partial}{\partial \rho}f_1(\rho,s)=s f_1(\rho,s)$.

Consider the low- and high-frequency limits of these expressions. The functions $f_1$ and $f_2$ both have the limits $1$ for $\omega\rightarrow 0$ and $0$ for $\omega \rightarrow \infty$. Thus, as expected $\delta T \rightarrow \delta Q / (V_0 c_p)$ in the low-frequency limit when heat has had time to distribute throughout the sphere. At high frequency the temperature amplitude becomes 
\begin{equation} \label{eq:cbh}
 \delta T(r) \rightarrow -\frac{1}{V_0c_l}a_l \delta Q=-\frac{1}{V_0}(\frac{1}{c_l}-\frac{1}{c_p}) \delta Q \quad  \textrm{for} \quad \omega \rightarrow \infty\,.
\end{equation}
If we for a moment consider the non-relaxing case where the specific heats are real, we see that the temperature amplitude is in counter-phase to the heat amplitude since $c_l<c_p$. For a propagating thermal wave it would not be surprising that temperatures at some distance -- e.g., at a half wavelength -- had opposite phases. However, Eq. (\ref{eq:cbh}) holds throughout the sphere and is not associated with the diffusive thermal wave. This will be even more clear when we consider the response to a heat step input later on. 
                                                                                                                                                                                                                                                                                                                                                                 
\begin{figure}[H]
\begin{center}
\includegraphics[width=8.6cm]{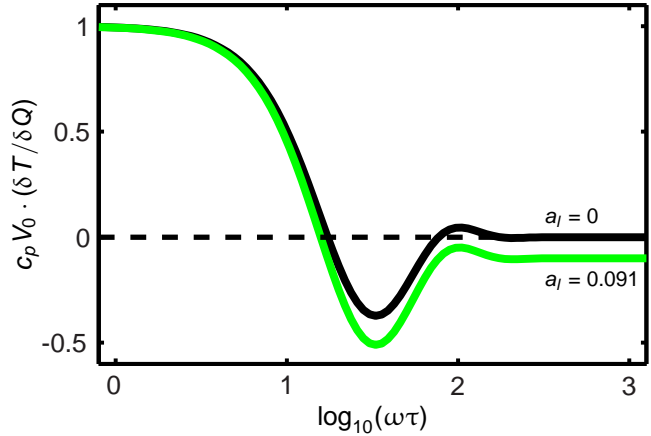}
\caption{The real part of the ratio between the complex
amplitudes of temperature at the center and the heat supplied at the surface scaled with the isobaric heat
capacity. At high frequencies the limit becomes the negative value  $-a_l/(1-a_l)=1-c_p/c_l$ .}\label{Afig1}
\end{center}
\end{figure}

We see that the longitudinal coupling constant $a_l$ controls the magnitude of the cooling-by-heating effect. The ratio of the amplitudes of the temperature in the center and the heat input at the surface is shown in Fig. \ref{Afig1}.  The phenomenon ``cooling by heating'' is indicated at high frequencies, albeit this is more conspicuous in the time domain.

For the displacement field we find for the low-frequency limit the natural result, 
\begin{equation}
 u(r) \rightarrow \frac{1}{3} \alpha_p \frac{\delta Q}{V_0 c_p} r   \quad  \textrm{for} \quad \omega \rightarrow 0
\end{equation}
determined by the final temperature rise $\frac{\delta Q}{V_0 c_p}$ and the linear thermal expansion $\frac{1}{3} \alpha_p$. However, at high frequencies we find
\begin{equation}
  u(r) \rightarrow \frac{1}{3} \alpha_p \frac{\delta Q}{V_0 c_p} \frac{4 G}{3 M_S}r  \quad \textrm{for} \quad \omega \rightarrow \infty\,.
\end{equation}
Notice that this displacement, which is responsible for the cooling-by-heating effect, is only present when $G\neq0$.
 
In the spherically symmetric case there are only two different components of the stress tensor, $\sigma_{rr}$ and $\sigma_{\theta\theta}=\sigma_{\phi\phi}$. It follows from Eqs. (\ref{eq:tve1}) and (\ref{eq:tve2}) and the fact that $\epsilon_{rr}=\frac{\partial u}{\partial r}$ and $\epsilon_{\theta\theta}=\epsilon_{\phi\phi}= \frac{u}{r}$ that $\sigma_{rr}=(K_T+\frac{4}{3}G)\frac{\partial u}{\partial r}+2(K_T-\frac{2}{3}G)\frac{u}{r}-\beta_V \delta T$, which by Eqs. (\ref{eq:solution_1}) and (\ref{eq:solution_2}) becomes
\begin{equation}\label{eq:solution_3}
\sigma _{rr} (r,\omega)=\frac{4}{3}\frac{G}{M_T}\frac{\beta_V}{V_0 c_l}\left\lbrace 1-f_2(r/R, i \omega \tau)\right \rbrace \delta Q\,.
\end{equation}
Likewise,  $\sigma_{\theta\theta}=(K_T-\frac{2}{3}G)\frac{\partial u}{\partial r}+2(K_T+\frac{1}{3}G)\frac{u}{r}-\beta_V \delta T$, which becomes
\begin{equation}\label{eq:solution_4}
\sigma_{\theta\theta} (r,\omega)=\frac{4}{3}\frac{G}{M_T}\frac{\beta_V}{V_0 c_l}\left\lbrace 
1-\frac{3}{2}f_1(r/R, i \omega \tau) + \frac{1}{2}f_2(r/R, i \omega \tau)\right\rbrace \delta Q \,.
\end{equation}

Thus at high frequencies there is an isotropic, uniform  tensile stress in the interior of the sphere of the magnitude
\begin{equation} \label{eq:sigma0}
  \sigma_{rr} \, ,\, \sigma_{\theta \theta} \, , \, \sigma_{\phi\phi} \rightarrow \frac{4}{3}\frac{G}{M_T}  \frac{\beta_V}{V_0 c_l}  \delta Q \quad \textrm{for} \quad \omega \rightarrow \infty
\end{equation}
On the other hand, all stresses vanish for $\omega \rightarrow 0$ (as expected).

In order to gain insight into the cooling-by-heating effect and show the significance of the longitudinal coupling constant $a_l$ we transform Eqs. (\ref{eq:solution_1}), (\ref{eq:solution_3}) and (\ref{eq:solution_4}) into the time domain, but only for a solid, i.e., in the case when all constitutive properties are frequency independent. If a delta function heat flux is applied at $t=0$, the heat supplied at the surface is a Heaviside step function, $\delta Q(R,t)=\delta Q_0 H(t)$; in this case calculating the inverse Laplace-Stieltjes transform leads to the following expressions for the temperature and stresses  as functions of time after $t=0$ (see Appendix \ref{App2}):
\begin{eqnarray}
\delta T (r,t)&=&\frac{1}{c_lV_0}\left\{-a_l+ F_1(r/R,t/\tau) \right\} \delta Q_0 \,,     \label{eq:sol_1_time}     \\
\sigma_{rr} (r,t)&=&\frac{4}{3}\frac{G}{M_T}\frac{\beta_V}{V_0 c_l}\left\{1- F_2(r/R,t/\tau)\right\} \delta Q_0  \label{eq:sol_3_time}\,,\\
\sigma_{\theta \theta} (r,t)&=&\frac{4}{3}\frac{G}{M_T}\frac{\beta_V}{V_0 c_l}\left\{1-\frac{3}{2} F_1(r/R,t/\tau)+\frac{1}{2} F_2(r/R,t/\tau)\right \} \delta Q_0  \label{eq:sol_4_time} \,,
\end{eqnarray}
where
\begin{eqnarray}
F_1(r/R,t/\tau)&=&1+ \frac{2}{3}\frac{R}{r}\sum_{n=1}^{\infty} 
\frac{\sin(\frac{r}{R}x_n)}{\sin(x_n)}e^{-x_n^2t/\tau} \\
F_2(r/R,t/\tau)&=&1+2 \left(\frac{R}{r}\right)^3\sum_{n=1}^{\infty}
\frac{\sin(\frac{r}{R}x_n)-\frac{r}{R}x_n\cos(\frac{r}{R}x_n)}{x_n^2\sin(x_n)}e^{-x_n^2t/\tau}\,.
\end{eqnarray}

Here $x_1<x_2<...$ are the positive roots of the transcendental equation $x=\tan(x)$. Note that $F_1 $ and $ F_2$ $\rightarrow 0$ for $t \rightarrow 0$  and $F_1 $ and $ F_2$ $\rightarrow 1$ for $t \rightarrow \infty$.  When $a_l=0$ there is no cooling-by-heating effect according to Eq. (\ref{eq:sol_1_time}). Furthermore, $a_l=0$ implies either $G=0$ or $\beta_V=0$  and there is no immediate expansion and no induced stresses.

\begin{figure}
\begin{center}
\includegraphics[width=8.6cm]{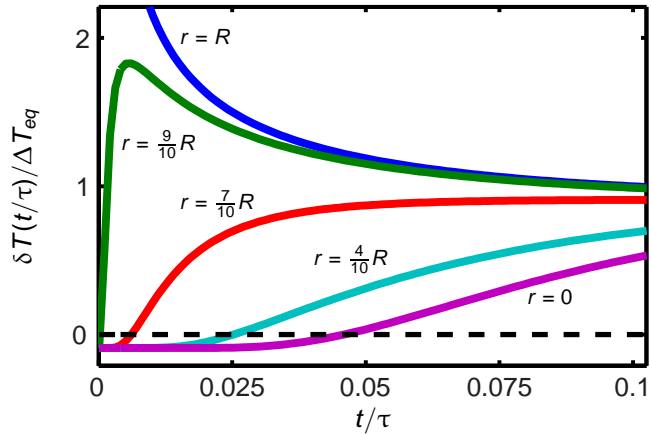}
\caption{\label{fig.2a} The temperature of the sphere as a function of time for different radii
normalized to the final temperature $\Delta T_{eq}=\delta Q_0/(V_0 c_p)$. After addition of heat at
the surface, the temperature drops instantaneously
throughout the sphere, showing adiabatic cooling. The time scale is given by the characteristic heat diffusion time, $\tau=R^2 c_l / \lambda$ . The longitudinal coupling constant is chosen to $a_l=0.091$.}
\end{center}
\end{figure}

When $a_l\neq 0$ the situation is quite different. In Fig. \ref{fig.2a} we plot the scaled temperature change $(c_pV_0 / \delta
Q)\delta T(t/\tau;r/R)$ for a several radii $r/R$ as given in Eq. (\ref{eq:sol_1_time}). The longitudinal coupling constant is here fixed to $a_l=0.091$ and time is given in units of the characteristic heat diffusion time $\tau$. The figure clearly shows the cooling-by-heating effect. Since a finite amount of heat was added at the surface at $t=0$, the surface temperature initially diverges. The interior of the sphere, even close to the surface, instantaneously cools to a uniform temperature. The expansion of the surface is immediately felt in the interior, and since no heat has yet arrived by diffusion, it cools adiabatically. This initial response is followed by an evolution in time where the temperatures of the different parts of the sphere converge and eventually equilibrate.

\begin{figure}
\begin{center}
\includegraphics[width=8.6cm]{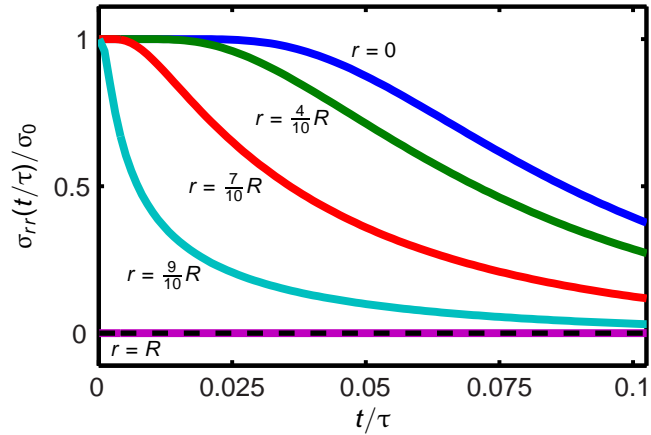}
\caption{The $rr$-component of the stress tensor, $1-F_2$, as a
function of time for a number of radii scaled with the initial stress, $\sigma_0=\frac{4}{3}\frac{G}{M_T}\frac{\beta_V}{V_0 c_l}\delta Q_0$. After the addition of heat at the surface, $\sigma_{rr}$ immediately increases throughout the sphere. The stress is released as heat diffuses towards the center from the surface.}\label{fig.2b}
\end{center}
\end{figure}

In order to understand better the physics of cooling-by-heating we consider the components of the
stresses given by Eqs. (\ref{eq:sol_3_time}) and (\ref{eq:sol_4_time}), respectively. In Fig.
\ref{fig.2b} the $\sigma_{rr}$ component of the stress
tensor is plotted scaled with the initial uniform interior stress $\sigma_0=\frac{4}{3}\frac{G}{M_T}\frac{\beta_V}{V_0 c_l}\delta Q_0$. First we note that the boundary condition $\sigma_{rr}(R,t)=0$ is fulfilled. As the surface receives heat and expands, an immediate traction is felt in the interior of the sphere. $\sigma_{rr}$ is positive, seeking to stretch a volume element in the radial direction under the entire evolution to thermal equilibrium.

\begin{figure}
\begin{center}
\includegraphics[width=8.6cm]{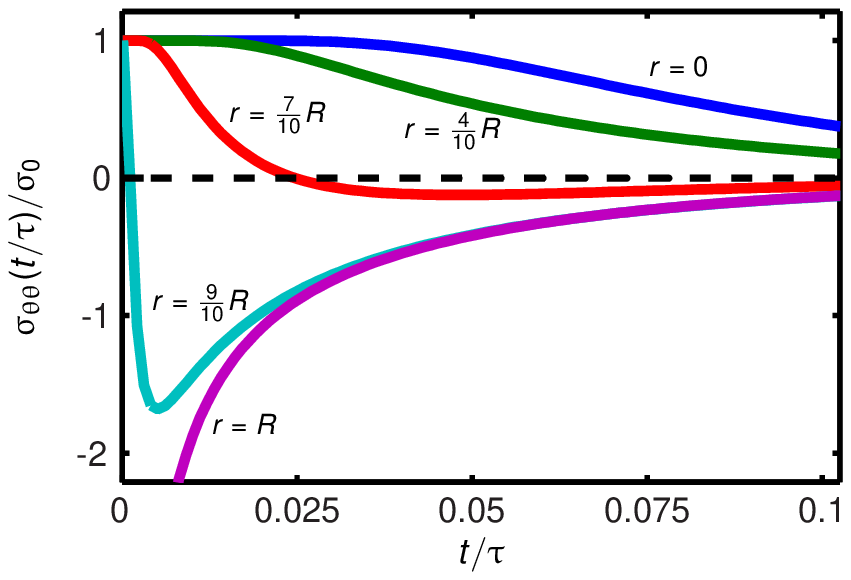}
\caption{The $\theta \theta$-component of the stress tensor, $1-\frac{3}{2}F_2+\frac{1}{2}F_1$, as a function of time scaled with the initial stress $\sigma_0$. There is an initial adiabatic positive step up in $\sigma_{\theta \theta}$  throughout the sphere. The regions that have been reached by the incoming diffusive heat at a certain point in time experience a negative $\theta\theta$-stress component since the inner unheated regions pull the outer heated regions inwards.}\label{fig.2c}
\end{center}
\end{figure}

The scaled stress component $\sigma_{\theta \theta} (r,t)$, is shown in Fig. \ref{fig.2c}. One
notices an immediate, uniform increase of this stress component throughout the sphere of the same
size as $\sigma_{rr}$. The initial stress is thus isotropic. Note that $\sigma_{\theta \theta}$
shifts sign during the thermal equilibration process, in contrast to $\sigma_{rr}$. This can be
understood in a physical picture: Consider the outer region that has been reached by the inflowing heat at
a certain point in time. If that region were free it would expand, but it is kept in place by the
inner unheated region that has not expanded thermally yet. This creates a negative stress on
surfaces with normal at right angle to the radius vector.

We conclude this section by a simple result. If one compares the instantaneous temperature drop Eq.
(\ref{eq:cbh}) and the instantaneous stress increase Eq. (\ref{eq:sigma0}), one finds that the
ratio is given by the adiabatic temperature---pressure coefficient:
\begin{equation}
 \frac{\delta T(r<R,t=0)}{\delta p(r<R,t=0)} = 1/\beta_S=\left(\frac{\partial T}{\partial
p}\right)_S\,.
\end{equation}

\subsection{The case when temperature is controlled at a mechanically free boundary}\label{sec2b}

The above studied case with heat-input control showed a rather simple cooling-by-heating behavior at short times or high frequencies. We now consider the case of controlling the temperature on the outer surface instead. There is still an effect, but it is not instantaneous. We only calculate the temperature in the center of the sphere. The surface is again mechanically free. Again, using the transfer matrix technique in the frequency domain, one finds that the temperature amplitude, $\delta T(0,s)$ in the center is related to the temperature amplitude, $\delta T(R,s)$ at the surface by $\delta T(0,s) =\Phi(s)\delta T(R,s)$, where
\begin{equation}
\Phi(s)= \left( 1-\frac{x^3-x^2\sinh(x)}{3a_l[ x\cosh(x)-\sinh(x)]-x^2\sinh(x)}\right)\,. \label{cp3}
\end{equation}
Here $x=\sqrt{s \tau(\omega)}$, $s=i\omega$. The characteristic diffusion time $\tau(\omega)$ (Eq. (\ref{eq:tau})) and thermomechanical coupling constant $a_l(\omega)$ (Eq. (\ref{a_l})) are in the general thermoviscoelastic case complex and frequency dependent and the temperature response can only be converted to the time domain numerically.

In order to calculate the temperature signal as a function of time we again limit ourselves to the purely thermoelastic case, i.e., the case of a solid where $\tau$ and $a_l$ are real and frequency independent. For a Heaviside temperature step at the surface of the sphere, $\delta T(R,t)=\Delta T H(t)$, the temperature at the sphere center is calculated via an inverse
Laplace-Stieltjes transform of $\Phi(s)$,
\begin{equation}\label{finalsolution}
 \delta T(0,t)=\Delta T \left\lbrace 1-
\displaystyle\sum_{k=0}^{\infty} R_k \exp \left(
-y_k^2\frac{t}{\tau}\right)      \right\rbrace \,,
\end{equation}
where the residues are given by 
\begin{equation}\label{residual}
 R_k=\frac{2(1-\cos{y_k}) +y_k\sin{y_k}
/(3a_l)}{(1-3a_l)\cos{y_k}+y_k(2-3a_l)\sin{y_k}}.
\end{equation}
Here the $y_k$´s denote the positive roots of the transcendental equation
\begin{equation}\label{transcendental}
\cot(y)=\frac{1}{y}-\frac{y}{3a_l}\,.
\end{equation}

\begin{figure}[H]
\begin{center}
\includegraphics[width=8.6cm]{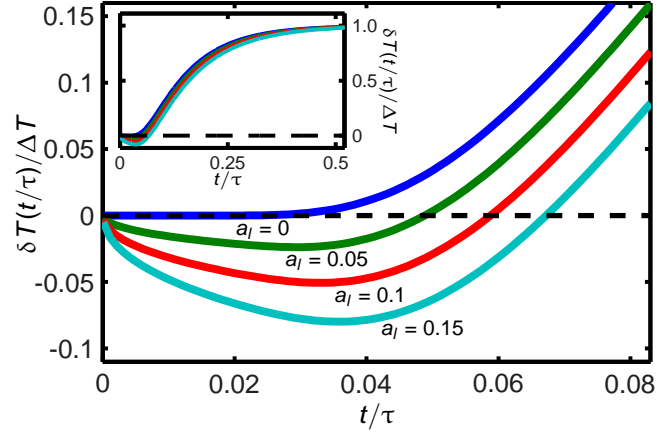}
\caption{The temperature at the center of the sphere as a function of time. After the temperature at the surface has been raised, the temperature at the center of the sphere initially decreases. This only happens when $c_l \neq c_p$, i.e., when the longitudinal coupling $a_l$ is not negligible. The temperature step at the boundary is $\Delta T=1$K, and the time scale is given by the characteristic diffusion time, $\tau=R^2 c_l / \lambda$.}\label{fig.3}
\end{center}
\end{figure}

In Fig. \ref{fig.3} we plot the solution Eq. (\ref{finalsolution}) for various values for the coupling constant $a_l$. Time is given in units of the characteristic diffusion time and $\Delta T=1$K. We see that the cooling-by-heating effect is present also when a step in temperature (instead of heat) is applied to the surface. However, now the effect is not instantaneous but evolves gradually, reflecting the gradual heat diffusion at the surface mediated to the center by the stress field. Figure (\ref{fig.3}) furthermore shows that it is not enough to have a thermomechanical coupling ($\alpha_p\neq 0$) for the phenomenon to be present -- only when $c_l \neq c_p$ is there a cooling-by-heating effect. The next section studies the general, thermoviscoelastic case of frequency  dependence of the response functions, which describes supercooled liquids.

\section{The thermoviscoelastic case}\label{sec4}

The above time-domain results apply for a thermoelastic solid only, whereas the frequency-domain
results are general. The thermoelastic examples handled so far only involved
frequency-\textit{independent} constitutive parameters, corresponding to the high-frequency
(low-temperature) limiting values of the curves sketched in Fig. \ref{fig1}. However, Fig. \ref{fig1} 
indicates that the value of the coupling constant $a_l$ is larger at lower frequencies
(higher temperature), thus suggesting that the effect of cooling by heating may be even larger in
the very viscous liquid or simply at the glass transition.  We investigate this issue in the time
domain in this section. In the thermoelastic case the inversion of the problem to the time-domain
could be made analytically. This is not possible in the thermoviscoelastic case where the
constitutive parameters are complex frequency-dependent functions.

\begin{figure}
\includegraphics[width=8.6cm]{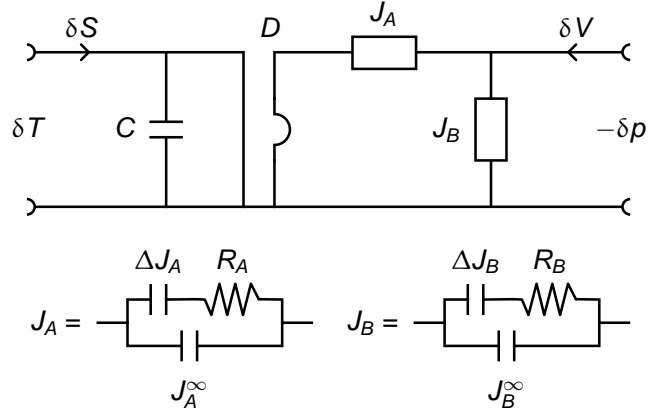}
\caption{\label{fig:dblrelnetwork} Electrical equivalent diagram of the interaction of a volume
element with its surroundings through two gates. The thermal gate where entropy
displacement $\delta S$ or temperature $\delta T$ can be controlled, and the
mechanical gate where volume displacement $\delta V$ or pressure $\delta p$ can be
controlled. $\delta S$ and $\delta V$ are generalized charge displacements and $\delta T$ and $-\delta p$
are generalized voltages. The relaxational elements $J_A$ and $J_B$ are simple single relaxation time elements.}
\end{figure}

In order to investigate the effect of going from solid to liquid we resort to numerical methods. Specifically, we transform $\Phi(s)$ of Eq. (\ref{cp3}) into the time-domain, accounting for the frequency-dependence of the constitutive parameters via $\tau$ and $a_l$. To do this we have to introduce a model of the constitutive parameters that enters via $\tau$ and $a_l$. It is common in rheology to illustrate models like the Maxwell model by rheological networks or even their electrical analogue. We use this approach to model the thermoviscoelastic behavior. The purpose of the model is to interpolate between the thermodynamic coefficients at high frequencies,  $\kappa_{T,\infty},c_{p,\infty},\alpha_{p,\infty}$ and at low frequencies  $\kappa_{T,0},c_{p,0},\alpha_{p,0}$. Network modeling assures internal consistency and agreement with the rules of linear irreversible thermodynamics. A one-parameter relaxation model implies the Prigogine-Defay ratio is unity, which is not the case for glucose. Rather, with $T_g=300$ K and $\Delta c_p=1.14 \cdot 10^6$ JK$^{-1}$m$^{-3}$, $\Delta \kappa_T=6.1 \cdot 10^{-11}$ Pa $^{-1}$ and $\Delta \alpha_p= 2.6 \cdot 10^{-4}$ K$^{-1}$ one finds
\begin{equation}
 \Pi=\frac{\Delta c_p \Delta \kappa_T}{T_g (\Delta \alpha_p)^2}=3.4\,.
\end{equation}
We are thus forced to consider a model with two relaxation elements that cannot be lumped into one. The model of Fig. \ref{fig:dblrelnetwork} is suited for this purpose. In order to still make it simple, the two relaxing elements are taken to be Debye-like. The model has a simple mathematical formulation in the frequency domain. We change the independent variables compared to Eqs. (\ref{eq:tve2}) and (\ref{eq:tve3}) and consider the complex amplitude $\delta T$ and $\delta p$ to be the controlled stimuli creating a linear response in the amplitudes $\delta \scrs$ and $\delta \epsilon$ of the entropy density and dilation: 
\begin{eqnarray}\label{compliancematrix}
\begin{pmatrix}
\delta \scrs \\ \delta \epsilon 
\end{pmatrix}
&=&  
\left(
  \begin{array}{ c c }
	\zeta_p(\omega)  & \alpha_p(\omega) \\
	\alpha_p(\omega) & \kappa_T(\omega) \\
  \end{array} 
\right)\cdot
\begin{pmatrix}
\delta T \\ -\delta p
\end{pmatrix} \,.
\end{eqnarray}
Here $\zeta_p=c_p/T_0$, $\alpha_p=\beta_V/K_T$, $\zeta_p=\zeta_V+\beta_V^2/K_T$ and $\kappa_T=1/K_T$.
In the model the three measurable quantities, the isobaric specific heat, the isobaric expansivity, and the isothermal compressibility are related to the elements $D,C,J_A(\omega)$, and $J_B(\omega)$ by
\begin{eqnarray}
 \zeta_p&=& D^2 J_A(\omega)+C \,,\\
 \alpha_p&=&-D J_A(\omega) \,,\\
 \kappa_T&=&J_A(\omega)+J_B(\omega) \,.
\end{eqnarray}
The relaxing element $J_A$ is determined by three parameters $J_{A,\infty}$, $\Delta J_A= J_{A,0}-J_{A,\infty}$ and $R_A$:
\begin{equation}
 J_A=J_{A,\infty}+\frac{1}{\frac{1}{\Delta J_A}+i\omega R_A}
\end{equation}
and likewise for $J_B$. 

The parameters of the model can be established from the high and low frequency limits of $\zeta_p$, $\kappa_T$ and $\alpha_p$. One finds $D=-\Delta \zeta_p / \Delta \alpha_p$, $J_{A,0}=-\alpha_{p,0}/D$, $J_{A,\infty}=-\alpha_{p,\infty}/D$, $J_{B,0}=\kappa_{T,0}-J_{A,0}$ and $J_{B,\infty}=\kappa_{T,\infty}-J_{A,\infty}$. The Prigogine-Defay ratio of the model is given by
\begin{equation}
 \Pi=\frac{\Delta \zeta_p \Delta \kappa_T}{(\Delta \alpha_p)^2}=1+\frac{\Delta J_B}{\Delta J_A}\,,
\end{equation}
and the dynamic (frequency-dependent) Prigogine-Defay ratio \cite{Ellegaard} is given by
 \begin{equation}
 \Lambda=\frac{\zeta_p'' \kappa_T''}{(\alpha_p'')^2}=1+\frac{ J_B''}{ J_A''}
\end{equation}

As expected, the Prigogine-Defay ratio is larger than unity, but becomes one if the element $J_B$ is non relaxing, in which case the model reduces to a single-parameter model. In the case of $J_A$ being non-relaxing, the model degenerates with absence of relaxation in $\zeta_p$ and $\alpha_p$. For glucose at $300$K we calculate the parameters of the model to be 
$D=-1.46 \cdot 10^7$ PaK$^{-1}$,
$C=4.76 \cdot 10^3$ JK$^{-1}$m$^{-3}$, 
$J_{A,\infty}=7.53 \cdot 10^{-12}$ Pa$^{-1}$, 
$\Delta J_A=1.78 \cdot 10^{-11}$ Pa$^{-1}$,
$J_{B,\infty}=8.55 \cdot 10^{-11}$ Pa$^{-1}$, and
$\Delta J_B=4.33 \cdot 10^{-11}$ Pa$^{-1}$.

The heat conductivity of glucose --- needed to calculate the heat diffusion time --- is $\lambda=0.35$ WK$^{-1}$m$^{-1}$ at $303$ K according to Greene and Parks \cite{GreeneParks1941}. The shear relaxation of glucose is modeled by a Maxwell model
\begin{equation}
 G(\omega)=\frac{1}{\frac{1}{G_\infty}+\frac{1}{i\omega\eta}}\,.
\end{equation}
Here the high-frequency shear modulus can be taken to be $G_\infty=3.1 \cdot 10^9$ Pa \cite{Meyer65}. The values of the thermodynamic parameters used to parametrize the model are given in Table \ref{table1}.
The temperature dependence of the shear viscosity causes the shift in the loss peaks of the relaxations. Parks {\it et al.} \cite{Parks1934} measured the viscosity of glucose in a wide temperature range from $295$K to $418$K. We fitted their tabulated data by the expression $\eta(T)=0.0125 \exp((512.9\textnormal{K}/T)^{6.42})$ Pas. This holds within $20\%$ over the entire temperature range except for the highest viscosity point of $9.1\cdot 10^{12}$ Pas, which however is well beyond the glass transition and may be hard to measure reliably. A Vogel-Fulcher law is not as good a fit, deviating more than $40\%$ in the measured temperature range. The rate parameters $R_A=R_A(T)$ and $R_B=R_B(T)$ are assumed to follow the temperature dependence of the viscosity \cite{Bo2012}. It is found numerically that one should chose $R_A(T)=R_B(T)=30\eta(T)$ in order to get the loss-peak frequency of the shear modulus, $G(\omega)$, and isothermal bulk modulus, $K_T(\omega)$, to coincide. The relaxation of the different response functions described by the model implies a frequency dependence of the longitudinal thermomechanical coupling via
\begin{equation}
 a_l(\omega)=\frac{\frac{4}{3}G(\omega)T_0\alpha_p^2(\omega)}{(1+\frac{4}{3}G(\omega)\kappa_T(\omega))c_p(\omega)}\,.
\end{equation}
The modulus of this complex function was shown in Figure (\ref{fig1}). Also, the heat-diffusion time 
\begin{equation}
 \tau(\omega)=R^2(1-a_l(\omega))c_p(\omega)/\lambda
\end{equation}
now becomes complex. This makes the inversion to the time-domain non-trivial and thus Eq. (\ref{cp3}) was inverted
numerically. The algorithm for the inverse Laplace transform is an improved version of de
Hoog's quotient difference method \cite{deHoog82} developed and implemented in Matlab by
Hollenbeck \cite{Hollenbeck98}.

 \begin{table}
 \ra{1.3}
 \begin{tabular}{@{}lrclrclr@{}}\toprule[0.3mm]
\multicolumn{2}{c}{Quantity} & \phantom{abc} & \multicolumn{2}{c}{Value} &\phantom{abc} &
\multicolumn{2}{c}{Reference} \\ \midrule[0.15mm]
$\omega\rightarrow 0$ \\
 &$\kappa_{T,0}$ & & & $15.4\cdot 10^{-11} \mathrm{Pa}^{-1}$ & & \cite{DaviesJones53}\\
 &$c_{p,0}$ & & & $3.05\cdot 10^{6} \mathrm{JK}^{-1}\mathrm{m}^{-3}$ & & \cite{Parks1928}\\
 &$\alpha_{p,0}$ & & & $3.7\cdot 10^{-4} \mathrm{K}^{-1}$ & & \cite{Parks1928}\\
&$\lambda$ (at $303\mathrm{K}$) & & & $0.35 \mathrm{WK}^{-1}\mathrm{m}^{-1}$ & & \cite{GreeneParks1941}\\
$\omega\rightarrow \infty$ \\
& $\kappa_{T,\infty}$ & & & $9.30\cdot 10^{-11} \mathrm{Pa}^{-1}$ & & \cite{DaviesJones53}\\
 &$c_{p,\infty}$ & & & $15.4\cdot 10^{-11} \mathrm{JK}^{-1}\mathrm{m}^{-3}$ & & \cite{Parks1928}\\
 &$\alpha_{p,\infty}$ & & & $1.1\cdot 10^{-4} \mathrm{K}^{-1}$ & & \cite{Parks1928}\\
&$G_{\infty}$ & & & $3.1 \cdot 10^9 \mathrm{Pa}$ & & \cite{Meyer65}\\
\bottomrule[0.3mm]
 \end{tabular}
 \caption{\label{table1} Literature data for glucose (at $300\mathrm{K}$) used to parametrize the model depicted in the electrical equivalent diagram in Fig. \ref{fig:dblrelnetwork}. We have used a density of $1.52 \cdot 10^3\mathrm{kg}\,\mathrm{m}^{-3}$ to convert specific heat data from mass to volume.}
 \end{table}

\begin{figure}
\includegraphics[width=8.6cm]{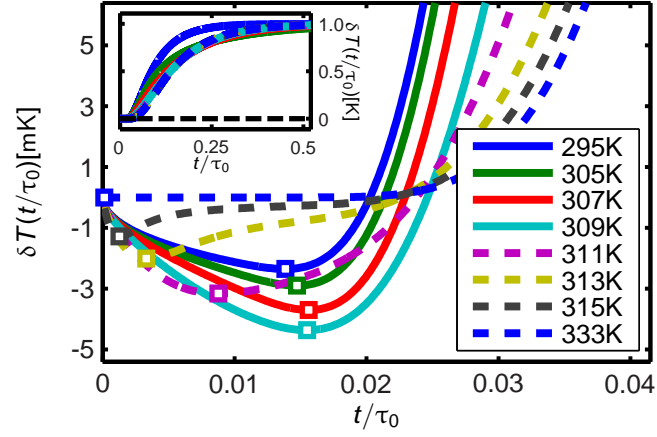}
\caption{\label{Figure8}The change in temperature, $\delta T$, at the center of a sphere, after a temperature step of $\Delta T=1 K$ has been applied at the surface. A model of the thermoviscoelastic relaxation of glucose with realistic limiting thermodynamic parameters has been invoked. Time is scaled by the characteristic diffusion time $\tau_0$, which is $800$ s for a glucose ball of radius $9.5$ mm. The minimum of approximately $-5$mK occurs for the $309$K curve just above $T_g$ at $0.017 \tau_0$, corresponding to $14$ s. Notice that although the cooling-by-heating phenomenon is present in the glassy solid state, it becomes more pronounced at the glass transition. The squares marking the minima at each temperature are plotted against temperature in Fig. \ref{Figure10}.}
\end{figure}

\begin{figure}
\includegraphics[width=8.6cm]{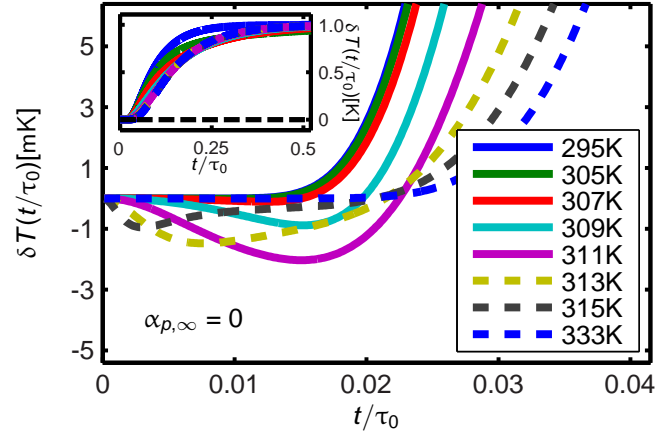}
\caption{\label{Figure9} A simulation similar to the one shown in Fig. (\ref{Figure8}) except that $\alpha_{p,\infty}$ has been put to $0$, forcing cooling by heating to be absent in the glassy solid phase. It is seen that the cooling by heating effect still appears as a dynamic phenomenon at the glass transition.}
\end{figure}

\begin{figure}
\includegraphics[width=8.6cm]{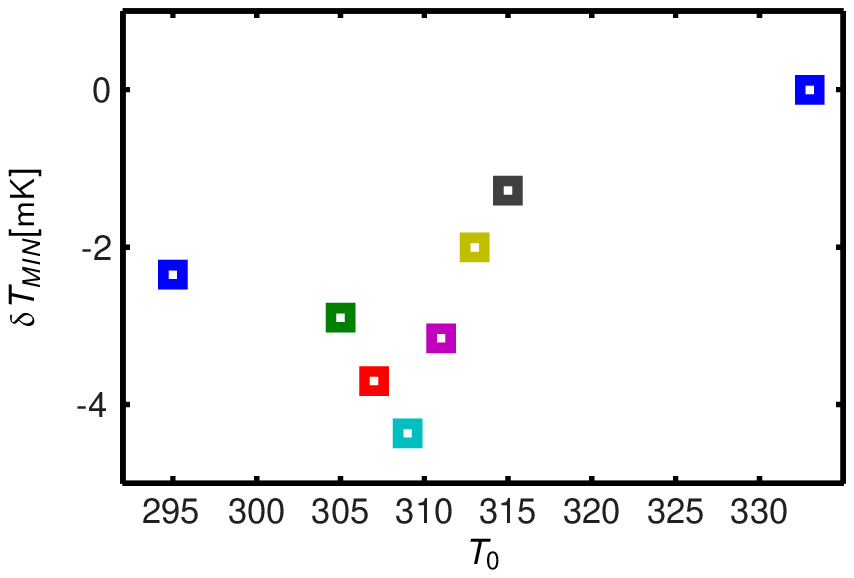}
\caption{\label{Figure10} The minima from Fig. \ref{Figure8} as a function of temperature. At high temperatures the phenomenon of cooling-by-heating is absent. Going down in temperature the liquid becomes more viscous and a dip in temperature emerges. Close to the glass transition the cooling-by-heating phenomenon gets most pronounced. As temperature is decreased further and the liquid enters the glassy state the effect is still present, however reduced in strength.}
\end{figure}

The calculated temperature response in the middle of the sphere to a step of $1$ K at the surface is shown in
Fig. \ref{Figure8}. Time is now scaled by the fixed real-valued diffusion time $\tau_0$ in the liquid regime,   
\begin{equation}
 \tau_0=R^2 c_{p,0}/\lambda\,.
\end{equation}
The figure shows that the effect of the thermomechanical coupling is absent at high temperatures.
But as temperature is decreased and the liquid gets more and more viscous, a dip in temperature
emerges. Going further down in temperature the phenomenon of cooling by heating becomes most
pronounced slightly above $T_g$. Even further down in temperature, in the glassy state, the effect
is still present, but small. One may ask what happens if the expansivity $\alpha_{p,\infty}$
vanishes, so that the phenomenon is absent in the glassy state: Will it still be present at the
glass transition? The simulations shown in Fig. (\ref{Figure9}) confirm this expectation. Putting 
$\alpha_{p,\infty}=0$, but otherwise keeping the values of the rest of the parameters, we get a
succession of temperature evolutions. Going down in temperature we see the cooling-by-heating
phenomenon appearing at $T_g$ and afterward disappearing in the glassy state. Fig. \ref{Figure10} shows the minimum temperature $\delta T_{\text{MIN}}$ reached when $\alpha_{p,\infty}\neq0$ as a function of temperature $T_0$, emphasizing the phenomenon
as characteristic of the glass transition.

\section{Experimental verification of the cooling-by-heating effect}\label{sec5}

\begin{figure}
\includegraphics[width=8.6cm]{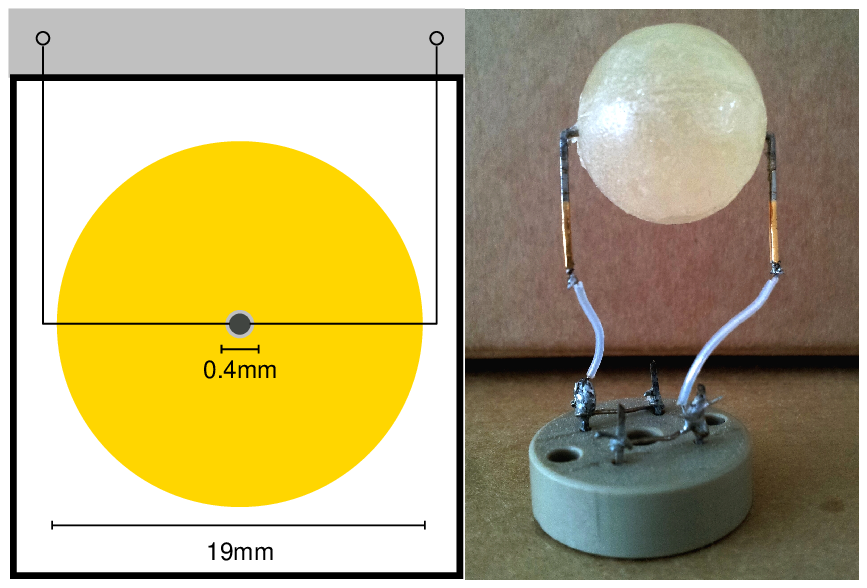}
\caption{\label{Figure11} Sketch of the experimental setup. The liquid is molded into a sphere, in which a small NTC-thermistor bead is placed in the center, connected to wires that lead to a multimeter that performs resistance measurements. The sphere is inserted into the cylindrical chamber of a cryostat. The photo shows one of the samples; at the time of the photo shoot (1 month after molding) the sample is no longer transparent because it crystallized.}
\end{figure}

To prove the existence of cooling by heating we molded glucose ($\alpha$-D(+) glucose, 98\%,
Sigma-Aldrich) into spherical samples with a thermistor placed in the center. Via the large negative
temperature coefficient (NTC) thermistor we measured the temperature in the middle of the sphere.
During measurements the glucose sphere is placed in a cryostat, that makes it possible to change the
temperature at the surface of the sphere quickly compared to the characteristic heat diffusion time. 
A sketch of the setup together with a photo of one of the samples is shown
in Fig. \ref{Figure11}. The photo shows the wires that lead into the sphere, connecting the thermistor
to the terminals on the peek plate shown in the photo. When mounted on a holder, the terminals get
connected to the multimeter that does the resistance measurement.

\begin{figure}
\begin{center}
\includegraphics[width=8.6cm]{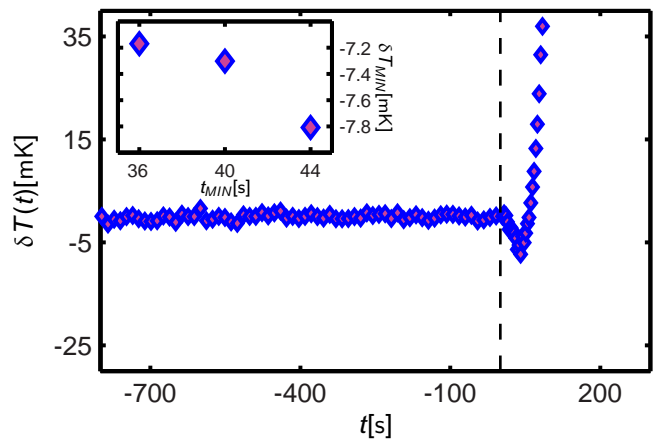}
\caption{\label{Figure12} The temperature in the center of a sphere of glucose, with diameter
$19\mathrm{mm}$. At time $t=0\mathrm{s}$ a step in temperature from $298\mathrm{K}$ to
$303\mathrm{K}$ is made with the cryostat (see the setup in Fig. \ref{Figure11}). The temperature
drops initially by $7.3\mathrm{mK}$ as result of cooling-by-heating. The inset shows the minimum
temperature $\delta T_{MIN}$ reached in three measurements, on three different samples, at the same
temperature.}
\end{center}
\end{figure}

The procedure in the experiments was the following. First we brought down the temperature to the desired starting level.
Then we waited for the temperature to equilibrate. This was monitored by measuring the resistance
every fifth minute. Typical waiting time was 18 hours. After the initial waiting time, we increased
the sampling rate of the multimeter to about five data points per minute. This was done for one hour
to get a baseline like the one in Fig. \ref{Figure12}. Then we imposed a $5\textrm{K}$ temperature
step and continued sampling data for another ten minutes with a sampling rate of 15 data points per
minute. Fig. \ref{Figure12} shows the temperature measured by the NTC thermistor during a
measurement with a step from $298\textrm{K}$ to $303\textrm{K}$ of the cryostat temperature. The
baseline extends for about $20$ minutes, then a characteristic temperature dip appears. The
magnitude of the dip is $7.3\pm0.2\textrm{mK}$, which was reached $40$ seconds
after the temperature step was imposed. 

The experiment was repeated on three different samples; the inset in Fig. \ref{Figure12} shows the results of the first measurement done on each sample, at the same temperature. Each marker represents the lowest temperature reached in one measurement. They are plotted against time after the temperature step is initiated. It was not possible to reproduce the phenomenon on the same sample by recycling the temperature. We ascribe this to crystallization. Although the effect should be present also in the solid state, it is here considerably  smaller and not observable with our temperature resolution. Nevertheless, the phenomenon was seen every time we repeated the experiment with a fresh supercooled sample. The sphere does not flow or deform to any appreciable degree even somewhat above the glass transition. The characteristic flow time $\tau_\text{flow}$ is proportional to the viscosity and inversely proportional to the gravitational force $mg$. By a dimensional argument it follows that
\begin{equation}
\tau_\text{flow} \, \propto \, \frac{\eta}{\rho g r} \,=\, \frac{G_\infty}{\rho g r} \tau_\text{M}\,,
\end{equation}
which means that $\tau_\text{flow} \sim 10^7 \tau_\text{M}$. Thus even at $310$ K, where  viscosity becomes $10^{10}$ Pas and thereby the Maxwell time $3$ s, the flow time is one year. 

\section{Discussion}
The concept of a longitudinal specific heat was identified in Ref. \onlinecite{Tage_1} as the
relevant quantity within AC-calorimetric methods that utilize heat effusion. The principle of the
simplest of these techniques \cite{BirgeNagel} is to measure the complex temperature response
$T_\omega$ at a plane surface to a heat current density $j_{Q,\omega}$ generated at the same
surface. The effusivity $e=\sqrt{\lambda c}$ is found from the measured specific thermal impedance
$Z \equiv T_\omega/j_{Q,\omega}=1/\sqrt{i\omega\lambda c}$, and from the effusivity the specific
heat can be calculated. A meticulous analysis of the thermomechanical equations of this problem showed that the
specific heat that comes into play in this situation is $c_l$ rather than $c_p$. Effusivity
measurements in spherical geometry have been shown also to involve the longitudinal specific heat
\cite{Tage_2,BoBoyeTage}. It is not a very well-known property, but it does appear in the textbook
on elasticity by Landau and Lifshitz \cite{Landau}. They show that the coupled thermoelastic
equations decouple for certain boundary conditions of an infinite solid, namely when temperature at
infinity is constant and deformation there is zero. They show further that the heat-diffusion
equation is valid with a diffusion constant containing the effective specific heat
$((1+\sigma)c_p+2((1-2\sigma)c_V)/(3(1-\sigma))$, where $\sigma$ is the isothermal Poisson ratio.
Inserting $\sigma=(3K_T-2G)/(6K_T+2G)$ one readily finds that the effective specific heat is $c_l$.
The longitudinal specific heat also appeared in Biot's 1956 paper \cite{Biot56} in his diffusion
equation for the entropy density. Although not very different from $c_p$, there is a fundamental
difference, and $c_l$ pops up in many thermoelastic problems when they are treated exactly. In
particular, as we have seen in this paper, there is only a cooling-by-heating effect if $c_l \neq
c_p$. We originally proposed the name longitudinal specific heat because this is the heat needed to
increase temperature by $1$ K if the associated expansion is confined to be longitudinal instead of
isotropic.
  
Transient thermal stresses induced by surface heating of a sphere have been considered theoretically by Cheung et al. \cite{Cheung74}. Their interest was fragmentation of brittle solids by surface heating. A heat current was applied uniformly within the polar angle regime $0 \leq \theta \leq \theta_0$ and the temperature and stress distributions calculated in time and space. This is a problem very similar to the one considered in this paper, although not generalized to situations with relaxation. However the cooling-by-heating phenomenon was not seen, since the standard decoupled heat-diffusion equation was used. The phenomenon thus seems apparently not to have been recognized in the literature, even though it belongs to classical continuum physics.

Other kinds of phenomena have lately been termed by the phrase ``cooling-by-heating'' or related names. Thus in 1999 Aleshin {\it et al.} reported ``heating through cooling'' when a copper bar is first heated to 150 $^o$C and subsequently rapidly cooled. This result in a temperature increase at the other end of the bar of about 4$^o$C \cite{ale99}. The authors presented the following explanation: The sudden cooling causes the bar to contract, producing an elastic wave that propagates towards the cold end of the bar. Under certain conditions this wave triggers a sequence of events responsible for an energy release, similar to a process called a ``steam explosion'' in an ionic liquid which is observed when a water jet interacts with a molten salt, an explosion that does not take place where the water hits the molten salt, but at the
bottom of the container \cite{ale97}. In 2007 Zwickl {\it et al.} reported a ``counter-intuitive cooling-by-heating'' effect for laser cooling of a microcantilever, an observation they suggested is due to photothermal forces causing the lowest cantilever vibrational mode to cool while all other modes are heated \cite{zwi07}. Very recently Mari and Eisert discussed theoretically cooling of quantum systems by means of incoherent thermal light \cite{mar11}. While coherent driving of a quantum system can mimic the effect of a cold thermal bath, the novel idea is that under certain conditions even  incoherent ``thermal'' light can be used to cool a quantum system.

\section{Summary}\label{sec6}

We have shown that cooling by heating occurs at the center of a solid spherical sample if it is heated at a mechanically free surface, reflecting a non-trivial thermomechanical coupling where the temperature initially decreases in the interior of the sphere. What happens is that, as heat diffuses into the outermost parts of the sphere, these parts expand and build up a negative pressure at the center of the sphere. This negative pressure couples to the temperature via the adiabatic pressure coefficient $\left(\frac{\partial T}{\partial p}\right)_S$. The opposite effect also applies, of course: if the temperature of the surface is lowered, heating by cooling will be observed. In ordinary solids the cooling-by-heating effect is almost negligible because their thermal expansion is generally small. The effect is particularly large for liquids close to their
glass transition. The cooling-by-heating phenomenon establishes the difference between the longitudinal and isobaric specific heat, since the effect is only present when these two quantities differ. This is the case when the shear modulus is non-vanishing compared to the bulk modulus and, simultaneously, the isobaric and isochoric specific heats differ  significantly. Analytical results show that the phenomenon occurs in the elastic case (the solid), and numerical results based on a model of the glass transition with parameters determined by glucose data show that the effect is present dynamically also in the very viscous liquid. Even in the hypothetical case when the glassy state is assumed to have zero expansivity, the phenomenon still appears at the glass transition.

The numerical results based on glucose data indicate that the drop in temperature at the center of the sphere is of order $5$ mK with a duration of approximately $15$ seconds when the temperature is increased by $1$K on the surface of a sphere of radius $r=10$ mm. This prediction was confirmed experimentally.

\acknowledgments
This work grew out of a suggestion by Niels Boye Olsen to whom we are obviously indebted. The center for viscous liquid
dynamics ``Glass and Time'' is sponsored by The Danish National Research Foundation.

\appendix
\section*{Appendix}
\setcounter{section}{1}
\subsection{Solution from section \ref{sec2} - the transfer matrix method}\label{App1}

The solution to the problem given in section \ref{sec2} is based on the transfer matrix formulation \cite{Tage_1,Tage_2,Carslaw} of the general solution of the thermoviscoelastic problem in a spherically symmetric case. Including the radial stress field and the time-integrated heat-current density that relate to the temperature and displacement fields, the authors of Refs. \onlinecite{Tage_1,Tage_2} end up with four coupled equations to solve. Laplace transforming the equations relating the four fields and solving the resulting inhomogeneous system of four ordinary differential equations, the result is in the general form of a transfer matrix $\tilde{\textbf{T}}(j,i)$ that links the dimensionless complex amplitudes of the fields at the boundary $r_i$ with those at $r_j$:
\begin{equation}
\begin{pmatrix}\delta \tilde{p_r} \cr \delta \tilde{T} \cr \delta \tilde{V} \cr \delta \tilde{S}
\cr\end{pmatrix}_{j} 
\,=\, 
\tilde{\textbf{T}}(j,i) 
\begin{pmatrix}\delta \tilde{p_r} \cr \delta \tilde{T} \cr \delta \tilde{V} \cr \delta \tilde{S}
\cr\end{pmatrix}_{i}
\,.
\end{equation}
Here $\delta \tilde{S}$, $\delta \tilde{V}$, $\delta \tilde{T}$,  and $\delta \tilde{p_r}$ are the complex amplitudes of entropy, volume, temperature, and the radial component of pressure ($\delta \tilde{p_r}=-\tilde{\sigma}_{rr}$), respectively. The elements of the transfer matrix are given in reference \cite{Tage_2}. From this general solution one can work out different cases, like the ones in Sec. \ref{sec2}. The boundary condition at $\tilde{r}_1=0$, giving net flux of heat through the center of the sphere, was set equal to zero, $\delta \tilde{S}_{1}=0$, since heat is supplied uniformly across the surface  $\tilde{r}_3$ giving a spherically symmetric case. For the same reason, $\delta \tilde{V}_{1}=0$. At the mechanically free outer boundary
$\tilde{r}_3$ the heat supplied, $\delta \tilde{S}_{3}$, is given and $\delta \tilde{p}_{r,3}=0$. Letting $\tilde{r}=\tilde{r}_2$ be an intermediate variable radius between $\tilde{r}_1$ and $\tilde{r}_3$ one has 
\begin{equation}\label{eq:A1}
\begin{pmatrix}\delta \tilde{p}_r \cr \delta \tilde{T} \cr \delta \tilde{V}
\cr \delta \tilde{S}
 \cr\end{pmatrix}_2 = 
\tilde{\textbf{T}}(2,1) \begin{pmatrix}\delta \tilde{p}_r \cr \delta \tilde{T} \cr 0
\cr 0
\cr\end{pmatrix}_{1}.
\end{equation}
Also
\begin{equation}\label{eq:A2}
\begin{pmatrix} 0 \cr \delta \tilde{T} \cr \delta \tilde{V} \cr \delta
\tilde{S} 
\cr\end{pmatrix}_3 = 
\tilde{\textbf{T}}(3,1)
\begin{pmatrix}\delta \tilde{p}_r \cr \delta \tilde{T} \cr 0 \cr 0
\cr\end{pmatrix}_{1}.
\end{equation}
This leads to $\delta \tilde{p}_{r,1}=-\frac{\tilde{T}_{12}(3,1)}{\tilde{T}_{11}(3,1)}\delta \tilde{T}_1$, or
$\delta
\tilde{T}_1=-\frac{\tilde{T}_{11}(3,1)}{\tilde{T}_{12}(3,1)}\delta \tilde{p}_{r,1}$ whereby $\delta
\tilde{S}_3=(\tilde{T}_{42}(3,1)-\frac{\tilde{T}_{41}(3,1)\tilde{T}_{12}(3,1)}{\tilde{T}_{11}(3,1)})\delta
\tilde{T}_1=(\tilde{T}_{41}(3,1)-\frac{\tilde{T}_{42}(3,1)\tilde{T}_{11}(3,1)}{\tilde{T}_{12}(3,1)})\delta \tilde{p}_{r,1}$.\\
This implies
\begin{equation}
\delta \tilde{T}_1=
\frac{-\tilde{T}_{11}(3,1)}{\tilde{T}_{41}(3,1)\tilde{T}_{12}(3,1)-\tilde{T}_{42}(3,1)\tilde{T}_{11}(3,1)})\delta \tilde{S}_3
\end{equation}
\begin{equation}
\delta \tilde{p}_{r,1}=
\frac{\tilde{T}_{12}(3,1)}{\tilde{T}_{41}(3,1)\tilde{T}_{12}(3,1)-\tilde{T}_{42}(3,1)\tilde{T}_{11}(3,1)})\delta \tilde{S}_3\,.
\end{equation}
Inserting this into Eq. (\ref{eq:A1}) one has 
\begin{equation}
\delta \tilde{p}_{r}(\tilde{r})=\delta \tilde{p}_{r,2}=
\frac{\tilde{T}_{11}(2,1)\tilde{T}_{12}(3,1)-\tilde{T}_{12}(2,1)\tilde{T}_{11}(3,1)}{\tilde{T}_{41}(3,1)\tilde{T}_{12}(3,1)-\tilde{T}_{
42}(3,1)\tilde{T}_{11}(3,1)})\delta \tilde{S}_3
\end{equation}
\begin{equation}
\delta \tilde{T}(\tilde{r})=\delta \tilde{T}_2=
\frac{\tilde{T}_{21}(2,1)\tilde{T}_{12}(3,1)-\tilde{T}_{22}(2,1)\tilde{T}_{11}(3,1)}{\tilde{T}_{41}(3,1)\tilde{T}_{12}(3,1)-\tilde{T}_{
42}(3,1)\tilde{T}_{11}(3,1)})\delta \tilde{S}_3
\end{equation}
\begin{equation}
\delta \tilde{V}(\tilde{r})=\delta \tilde{V}_2=
\frac{\tilde{T}_{31}(2,1)\tilde{T}_{12}(3,1)-\tilde{T}_{32}(2,1)\tilde{T}_{11}(3,1)}{\tilde{T}_{41}(3,1)\tilde{T}_{12}(3,1)-\tilde{T}_{
42}(3,1)\tilde{T}_{11}(3,1)})\delta \tilde{S}_3
\end{equation}
\begin{equation}
\delta \tilde{S}(\tilde{r})=\delta \tilde{S}_2=
\frac{\tilde{T}_{41}(2,1)\tilde{T}_{12}(3,1)-\tilde{T}_{42}(2,1)\tilde{T}_{11}(3,1)}{\tilde{T}_{41}(3,1)\tilde{T}_{12}(3,1)-\tilde{T}_{
42}(3,1)\tilde{T}_{11}(3,1)})\delta \tilde{S}_3\,.
\end{equation}
Inserting the actual explicit values of the transfer matrix elements from \cite{Tage_2} and evaluating them in the limit of $\tilde{r}_1\rightarrow0$, putting $\tilde{r}=\tilde{r}_2$ $\tilde{R}=\tilde{r}_3$, yields
\begin{eqnarray}
 \delta \tilde{p}_{r}(\tilde{r})&=&\frac{3 \tilde{\alpha} \tilde{g}}{\tilde{c}(1+\tilde{g})} \left[ \frac{1}{\tilde{R}^3}  - \frac{\tilde{r} \cosh(\tilde{r})-\sinh(\tilde{r})}{\tilde{r}^3 (\tilde{R} \cosh(\tilde{R})-\sinh(\tilde{R}))}\right]\delta \tilde{S}(\tilde{R})\\
\delta \tilde{T}(\tilde{r})&=&\frac{1}{\tilde{c}} \left[ \frac{1}{\tilde{R}^3}\frac{3\tilde{\alpha}^2 \tilde{g}}{\tilde{\alpha}^2 \tilde{g} + \tilde{c}(1+\tilde{g})} - \frac{\sinh(\tilde{r})}{\tilde{r}(\tilde{R} \cosh(\tilde{R})-\sinh(\tilde{R}))}\right]\delta \tilde{S}(\tilde{R}) \label{eq:solution_1dl}\\
 \delta \tilde{V}(\tilde{r})&=&\frac{\tilde{\alpha}}{\tilde{c}(1+\tilde{g})} \left[ {\biggl(\frac{\tilde{r}}{\tilde{R}}\biggr)}^3\frac{\tilde{g}(\tilde{\alpha}^2 -\tilde{c}(1+\tilde{g}))}{\tilde{\alpha}^2 \tilde{g}+\tilde{c}(1+\tilde{g})}  - \frac{\tilde{r} \cosh(\tilde{r})-\sinh(\tilde{r})}{\tilde{R} \cosh(\tilde{R})-\sinh(\tilde{R})}\right]\delta \tilde{S}(\tilde{R})  \label{eq:solution_2dl} \\
 \delta \tilde{S}(\tilde{r})&=& \frac{\tilde{r} \cosh(\tilde{r})-\sinh(\tilde{r})}{\tilde{R} \cosh(\tilde{R})-\sinh(\tilde{R})}\delta \tilde{S}(\tilde{R})\,.
\end{eqnarray}

The transformation back to dimensionalized physical quantities is performed by noticing that $\tilde{r}=k r$,$\tilde{R}=k R$, $\tilde{u}=k u$, where $k=\sqrt{s c_l/\lambda}$. Furthermore $\tilde{c}=T_0 c_l/K_T$, $\tilde{\alpha}=T_0\alpha_p$, $\tilde{g}=4G/(3 K_T)$,  $\tilde{\sigma}_{rr}=\sigma_{rr}/K_T$, $\delta \tilde{V} = \delta V k^3/(4 \pi)$, $\delta S = \delta \tilde{S} k^3/(4 \pi K_T)$, $\tilde{\delta T}=\delta T /T_0$. Since the entropy-displacement is positive in the direction of $r$, it is related to the heat input at the outer surface by  $\delta Q = -T_0 \delta S$, opposite of the convention in Ref. \onlinecite{Tage_2}. Recalling that $\tilde{u}=\tilde{r}^2\tilde{V}$ one now easily derives Eqs. (\ref{eq:solution_1}) and (\ref{eq:solution_2}) from Eqs. (\ref{eq:solution_1dl}) and (\ref{eq:solution_2dl}).

\subsection{Inverse Laplace-Stieltjes transforms.}\label{App2}

If a stimulus $\phi_s e^{st}$ ($s=i\omega$) on a linear system gives rise to a response $\gamma_s e^{st}$, where $\gamma_s=f(s)\phi_s$, the response to a Heaviside input $\phi_0 H(t)$ will be $\gamma(t)=F(t)\phi_0$, where $F(t)$ is the inverse Laplace transform of $f(s)/s$  (or the inverse Laplace-Stieltjes transform of $f(s)$). We perform the inverse Laplace transform via the calculus of residues as
\begin{equation}
 F(t)=\sum_{\text{poles } s_n} \text{Res}(\frac{f(s)}{s},s_n) e^{s_n t}
\end{equation}
Put $\rho=r/R$ and $\tau=(c_l/\lambda) R^2$. Then $kr=\sqrt{s\tau}\rho$. Choose furthermore initially time units such that $\tau=1$. Then the two expressions of Eqs. (\ref{eq:f1}) and (\ref{eq:f2}) when divided by $s$ become
\newcommand{\sqrs}{\sqrt{s}}
\begin{eqnarray}
 \frac{f_1(s)}{s}&=&\frac{\sinh(\rho\sqrs)}{3\rho \sqrs} \frac{\sqrs}{\sqrs \cosh(\sqrs)-\sinh(\sqrs)} \\
 \frac{f_2(s)}{s}&=&\frac{1}{\rho^3 s} \frac{\rho \sqrs \cosh(\rho\sqrs)-\sinh(\rho \sqrs)}{\sqrs \cosh(\sqrs)-\sinh(\sqrs)}\,.
\end{eqnarray}
Both expressions have simple poles at $s=0$ with residue $1$. The other poles are on the negative real axis $s_n=-x_n^2$, where $x_n$ are the positive roots of the transcendental equation $\tan(x)=x$. $x_1 \approx 4.493409457$ and $x_2\approx7.725251836$ and $x_n \approx \sqrt{(\pi/2+n \pi)^2-2}$ better than $1$ ppm for $n\geq3$ . The residues now become, respectively
\begin{eqnarray}
 \text{Res}(\frac{f_1(s)}{s},-x_n^2)&=& \frac{2\sin(\rho x_n)}{3\rho \sin(x_n) } \\
 \text{Res}(\frac{f_2(s)}{s},-x_n^2)&=& \frac{2\sin(\rho x_n)-\rho x_n \cos(\rho x_n)}{\rho^3 x_n^2 \sin(x_n)}\,,
 \end{eqnarray}
and the corresponding time-domain functions 
\begin{eqnarray}
F_1(\rho,t)&=&1+ \frac{2}{3}\frac{1}{\rho}\sum_{n=1}^{\infty} 
\frac{\sin(\rho x_n)}{\sin(x_n)}e^{-x_n^2t} \\
F_2(\rho,t)&=&1+2 \left(\frac{1}{\rho}\right)^3\sum_{n=1}^{\infty}
\frac{\sin(\rho x_n)-\rho x_n\cos(\rho x_n)}{x_n^2\sin(x_n)}e^{-x_n^2t}
\end{eqnarray}
Using these expressions and reintroducing the characteristic heat diffusion time $\tau$, one derives Eqs. (\ref{eq:sol_1_time}), (\ref{eq:sol_3_time}), and (\ref{eq:sol_4_time}).

\end{document}